%% file: main.tex
\documentclass[11pt,letterpaper]{article}
\usepackage[a4paper,
            bindingoffset=0.2in,
            left=1in,
            right=1in,
            top=1in,
            bottom=1in,
            footskip=.25in]{geometry}
\usepackage{tabularx,booktabs}
\usepackage{url}
\usepackage{times}
\usepackage{fullpage}
\usepackage{graphicx}
\usepackage{lipsum}
\newcommand{\junk}[1]{}
\usepackage{longtable}
\usepackage{soul}
\usepackage{tabularx,booktabs}
\usepackage{multirow}
\usepackage{booktabs}
\usepackage{color}
\usepackage{subfigure}
\usepackage{amsfonts}
\usepackage{amsmath}
\usepackage[dvipsnames]{xcolor}
\usepackage{tikz}
\usepackage{threeparttable}
\usepackage{cite}
\usepackage{authblk}
\usepackage{caption}
\usepackage{lineno}
\usepackage[normalem]{ulem}
\linenumbers

\usepackage[title]{appendix}
\usepackage{supertabular}

\makeatletter
\newbox\dottedarrow@box
\setbox\dottedarrow@box\hbox
  {%
    \begin{tikzpicture}
      \draw[dotted,->] (0,0) -- (1.5em,0);
    \end{tikzpicture}%
  }
\newcommand*\dottedarrow
  {\relax\ifmmode\expandafter\dottedarrow@m\else\expandafter\dottedarrow@t\fi}
\newcommand*\dottedarrow@t[1][1.5em]
  {\resizebox{#1}{!}{\raisebox{.5ex}{\usebox\dottedarrow@box}}}
\newcommand*\dottedarrow@m[1][]
  {%
    \if\relax\detokenize{#1}\relax
      \mathchoice% values are trial and error based\ldots
        {\dottedarrow@t}
        {\dottedarrow@t}
        {\dottedarrow@t[1.1em]}
        {\dottedarrow@t[0.9em]}%
    \else
      \dottedarrow@t[#1]%
    \fi
  }
\makeatother

\newcommand{\highlight}[1]{\textcolor{black}{#1}}

\usepackage[numbers,sort&compress]{natbib}

\usepackage{natbib}
\setcitestyle{super}
\usepackage{hyperref}
\usepackage{xurl}

\begin{document}
\thispagestyle{empty}

\nolinenumbers
% \title{E2Efold-3D: End-to-End Deep Learning Method for accurate \textit{de novo} RNA 3D Structure Prediction}
\title{Accurate RNA 3D structure prediction using a language model-based deep learning approach}

\author[1,2,13]{Tao Shen$^{*}$}
\author[1]{Zhihang Hu$^{*}$}
\author[3,11]{Siqi Sun$^{*\dagger}$}
\author[4,5,14]{Di Liu$^{*\dagger}$}
\author[6,7,8,9]{Felix Wong}
\author[1,10]{Jiuming Wang}
\author[1]{Jiayang Chen}
\author[1]{Yixuan Wang}
\author[1]{Liang Hong}
\author[1]{Jin Xiao}
\author[2,13]{Liangzhen Zheng}
\author[15]{Tejas Krishnamoorthi}
\author[1]{Irwin King}
\author[2,13]{Sheng Wang$^{\dagger}$}
\author[4,5]{Peng Yin$^{\dagger}$}
\author[4,6,7,8]{James J. Collins$^{\dagger}$}
\author[1,4,6,7,12]{Yu Li$^{*\dagger}$}

\affil[1]{Department of Computer Science and Engineering, The Chinese University of Hong Kong, Hong Kong SAR, China}
\affil[2]{Zelixir Biotech, Shanghai, China}
\affil[3]{Research Institute of Intelligent Complex Systems, Fudan University, Shanghai, China}
\affil[4]{Wyss Institute for Biologically Inspired Engineering, Harvard University, Boston, MA, USA}
\affil[5]{Department of Systems Biology, Harvard Medical School, Boston, MA, USA}  
\affil[6]{Institute for Medical Engineering and Science, Massachusetts Institute of Technology, Cambridge, MA, USA}
\affil[7]{Broad Institute of MIT and Harvard, Cambridge, MA, USA}
\affil[8]{Synthetic Biology Center, Massachusetts Institute of Technology, Cambridge, MA, USA}
\affil[9]{Integrated Biosciences, Redwood City, CA, USA}
\affil[10]{OneAIM Ltd., Hong Kong SAR, China}
\affil[11]{Shanghai AI Laboratory, Shanghai, China}
\affil[12]{The CUHK Shenzhen Research Institute, Hi-Tech Park, Nanshan, Shenzhen, China}
\affil[13]{Shenzhen Institute of Advanced Technology, Shenzhen, Guangdong, China}
\affil[14]{School of Molecular Sciences and Center for Molecular Design and Biomimetics at the Biodesign Institute, Arizona State University, Tempe, AZ, USA}
\affil[15]{School of Computing and Augmented Intelligence, Arizona State University, Tempe, AZ, USA}

\affil[*]{\small These authors contributed equally to this work}
\affil[$\dagger$]{\small Corresponding authors: siqisun@fudan.edu.cn, wangsheng@zelixir.com, peng\_yin@hms.harvard.edu, jimjc@mit.edu, liyu@cse.cuhk.edu.hk}

\date{}

\maketitle
\begin{abstract}
  \input{abstract1.tex}
\end{abstract}

\section*{Introduction}
  \input{intro1.tex}

\section*{Results}
\label{result}
  \input{result_ash.tex}

\section*{Discussion}
\label{discussion}
  \input{discussion.tex}

\section*{Methods}
\label{method}
  \input{methods.tex}

\section*{Data Availability}
  \input{data.tex}
  
\section*{Code Availability}
  \input{code.tex}

\section*{Acknowledgments}
\input{acknowledgment.tex}

% \section{Acknowledgements} 
% This work was supported by a grant by ...

% \bibliographystyle{plainnat}
\bibliographystyle{unsrtnat}  % Jiuming: sort by first occurrence

\bibliography{rhofold_1101}
\pagebreak

% \begin{appendices}
% \section{Tables}
% \label{Appendix}
%   \input{appendix.tex}
% \end{appendices}

\end{document}

%% file: abstract1.tex
Accurate prediction of RNA three-dimensional (3D) structure remains an unsolved challenge. Determining RNA 3D structures is crucial for understanding their functions and informing RNA-targeting drug development and synthetic biology design. The structural flexibility of RNA, which leads to scarcity of experimentally determined data, complicates computational prediction efforts. Here, we present RhoFold+, an RNA language model-based deep learning method that accurately predicts 3D structures of single-chain RNAs from sequences. By integrating an RNA language model pre-trained on $\sim$23.7 million RNA sequences and leveraging techniques to address data scarcity, RhoFold+ offers a fully automated end-to-end pipeline for RNA 3D structure prediction. Retrospective evaluations on RNA-Puzzles and CASP15 natural RNA targets demonstrate RhoFold+'s superiority over existing methods, including human expert groups. Its efficacy and generalizability are further validated through cross-family and cross-type assessments, as well as time-censored benchmarks. Additionally, RhoFold+ predicts RNA secondary structures and inter-helical angles, providing empirically verifiable features that broaden its applicability to RNA structure and function studies.

%% file: intro1.tex
RNA molecules occupy a key role in the central dogma of molecular biology. How RNA structures impinge on gene regulation and function has been a subject of intense study \citep{mortimer2014insights}. Studies focusing on RNA targeting have demonstrated that it can be an important, druggable target for drug development \citep{warner2018principles,kulkarni2021current,sheridan2021first} and a useful synthetic biology design element \citep{zhao2022rna}.  Over 85\% of the human genome is transcribed, but a mere 3\% encodes proteins, underscoring the substantial portion of transcribed RNAs with unknown functions and structures.
In many cases, obtaining high-resolution structural information can enable a more predictive understanding of the RNA molecules of interest \citep{liu2022sub,sheridan2021first}.

The conformational flexibility of RNA molecules has made the experimental determination of their 3D structures challenging. As of December 2023, RNA-only structures comprise less than 1.0\% of the $\sim$214,000 structures in the Protein Data Bank (PDB), and RNA-containing complexes account for only 2.1\% \citep{xu2022recent,liu2022sub}. Despite advances in X-ray crystallography, NMR spectroscopy, and cryo-EM, these low-throughput techniques are limited by specialized requirements. Computational methods have emerged as a complementary approach for RNA 3D structure prediction, leveraging RNA sequence data. These methods fall into two main categories: template-based modeling, such as ModeRNA \citep{rother2011moderna} and RNAbuilder \citep{flores2010predicting}, which are constrained by limited template libraries; and \textit{de novo} prediction approaches, including FARFAR2 \citep{watkins2020farfar2}, 3dRNA \citep{wang20193drna}, and SimRNA \citep{boniecki2016simrna}, which are more predictive but computationally intensive due to large-scale sampling requirements.

An orthogonal \textit{de novo} prediction approach is to leverage deep learning, which has been successfully applied to various biological problems. These applications include predicting protein 3D structures \citep{af2}, RNA secondary structures \citep{chen2020rna, chen2022interpretable}, and scoring RNA structures generated by other methods \citep{townshend2021geometric}. Previous methods for RNA 3D structure prediction focused on template-based or energy-based sampling techniques, which were informed by the scarcity of available RNA 3D structural data. 
\highlight{Despite the scarcity of data, the success of AlphaFold2 \cite{af2} for protein structure prediction has catalyzed the development of \textit{de novo} deep learning methods for RNA 3D structure prediction. These \textit{de novo} methods often begin with a single input sequence and then construct multiple sequence alignments (MSAs) from it, which are subsequently used to build the 3D structures.}

MSAs have been shown to provide additional information helpful for protein modeling, and this may be similarly true for RNAs.  For instance, DeepFoldRNA \cite{pearce2022novo} and trRosettaRNA \cite{wang2023trrosettarna} utilize transformer networks (e.g., RNAformer) to convert built MSAs and predicted secondary structures into various 1D and 2D distances, orientations, and torsion angles. These predicted geometries are then leveraged as constraints to predict RNA 3D structures using energy minimization, integrating sampling and scoring processes into their frameworks. Several models, including E2Efold-3D \cite{shen2022e2efold} and RoseTTAFoldNA \cite{baek2023accurate},  employ fully differentiable end-to-end pipelines that directly predict all-atom 3D models using built MSAs and secondary structure constraints. 
\highlight{AlphaFold3~\cite{abramson2024accurate}, the successor to AlphaFold2~\cite{jumper2021highly}, is also capable of predicting RNA 3D structures directly from input sequences, while still relying on its constructed MSAs during the prediction process.} 
In contrast to other methods, AlphaFold3 \cite{abramson2024accurate} employs a diffusion-based process to predict raw atom coordinates, replacing the AlphaFold2 structure module operating on amino-acid-specific frames and side-chain torsion angles. While these MSA-based methods are capable of accurately predicting RNA 3D structures, they require extensive searches across large sequence databases, which can be time-consuming. In contrast, models based on single sequences, including DRFold \cite{li2023integrating}, do not utilize MSAs and thus do not require extensive searches in large sequence databases. Instead, DRFold \cite{li2023integrating} relies solely on predicted secondary structures to inform 3D structure predictions. This approach is faster, but typically has a lower accuracy compared to MSA-based methods. Next-generation deep learning methods might better leverage MSA-based approaches in a way that improves both speed and accuracy.

Here, we present a language model-based deep learning method, RhoFold+, for accurate and fast \textit{de novo} RNA three-dimensional structure prediction. RhoFold+ represents a fully automated and differentiable improvement over its predecessor, RhoFold \cite{shen2022e2efold}, leveraging improved integration of MSAs and other features to enhance performance. Our primary focus is on determining the structures of single-chain RNAs, which have limited interactions with other molecules. Addressing this challenge can help us better understand RNA biology and provide a starting point for solving more complex structural problems.

%% file: result_ash.tex
\begin{figure*}[!t]
    \centering
    \includegraphics[width=0.90\textwidth]{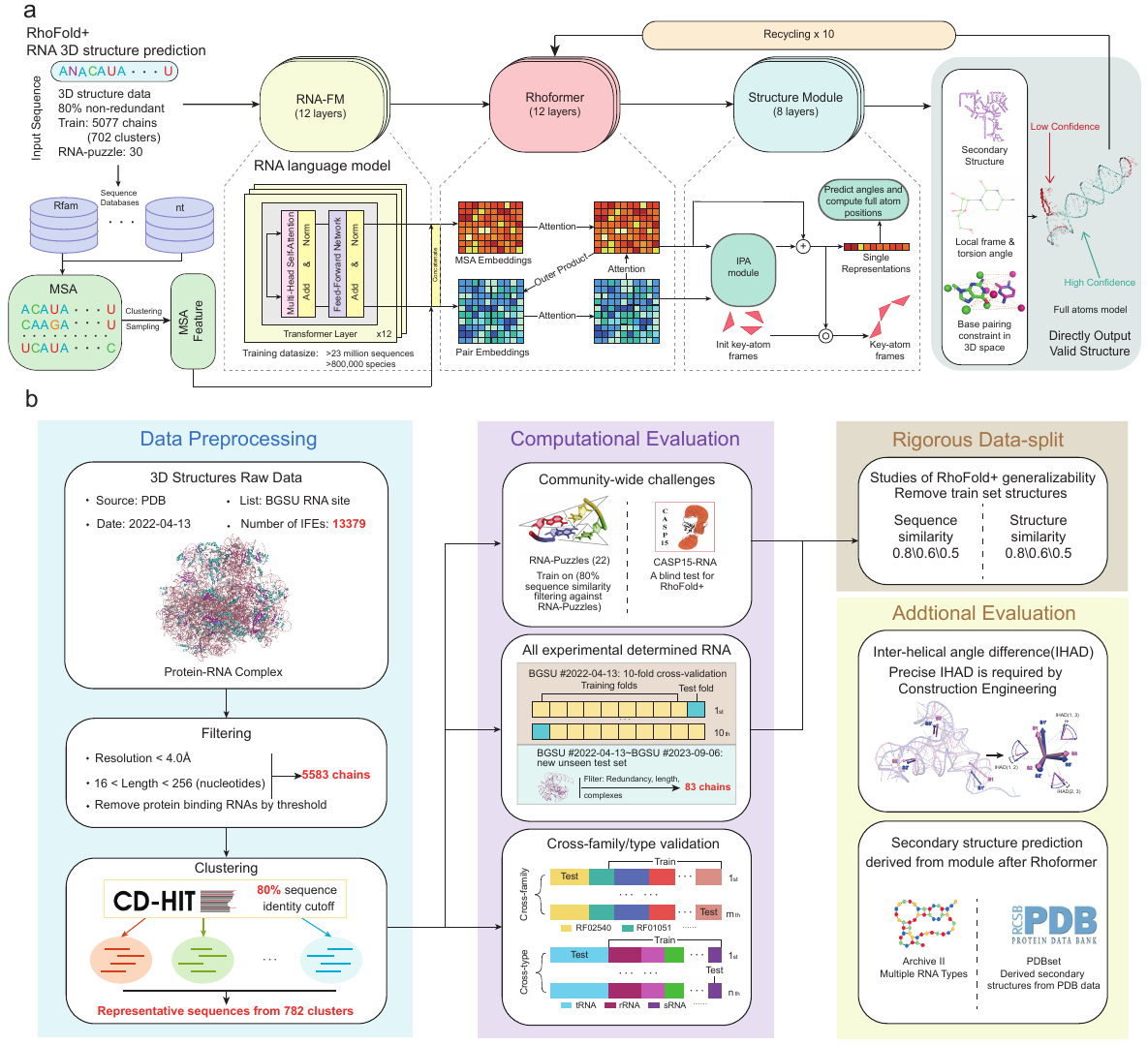}
    \caption{\textbf{The architecture of RhoFold+ and the tasks used for performance evaluation}. 
    \textbf{a.} The architecture of RhoFold+, a fully automated and differentiable end-to-end approach to \textit{de novo} RNA 3D structure prediction from sequence. Using an RNA language model (RNA-FM) pre-trained on 23,735,169 unannotated RNA sequences, and several deep learning modules---including an invariant point attention (IPA) module which models 3D positions---RhoFold+ can generate valid and largely accurate RNA 3D structures of interest typically within $\sim$0.14 seconds (w/o MSA searching). 
    \textbf{b.} The preprocessing step of RhoFold+ to extract all available non-redundant single-stranded RNA 3D structures from the PDB database. RhoFold+ is comprehensively benchmarked on community-wide challenges including RNA-Puzzles targets and CASP15 natural RNA targets, and on all available experimentally determined RNA 3D structures. RhoFold+ also demonstrates high accuracy in cross-validation experiments, as well as generalizability to unseen, newly determined RNA structures and unseen RNA families and types in cross-family and cross-type validation experiments. Data-split evaluations reveal that RhoFold+ does not overfit its training set. RhoFold+ is also capable of predicting secondary structures and parameters that are useful for construct engineering.}
    \label{fig:overview}
\end{figure*}
\paragraph{Automated end-to-end platform for RNA 3D structure prediction} 
The development of RhoFold+ was guided by RNA-specific knowledge and the limitations of existing RNA 3D structure data. To build our training dataset, we curated all available RNA 3D structures from the PDB, using the BGSU Representative Sets of RNA structures (version 2022-04-13). We focused on single-chain RNAs and reduced redundancy by clustering sequences with CD-HIT \cite{li2006cd} at an 80\% sequence similarity threshold, resulting in 782 unique sequence clusters from 5,583 RNA chains. These RNA sequences were then processed through our pipeline, RhoFold+. First, the sequences were transformed using RNA-FM, our large RNA language model, to extract evolutionarily and structurally informed embeddings. Concurrently, MSAs were generated by searching through extensive sequence databases. The embeddings and MSA features were then fed into our transformer network, Rhoformer, and iteratively refined for 10 cycles. Following this, our structure module employed a geometry-aware attention mechanism and an invariant point attention (IPA) module to optimize local frame coordinates and torsion angles for key atoms in the RNA backbone. Structural constraints, such as secondary structure and base pairing, were applied after reconstructing the full-atom coordinates (Fig.1a; detailed discussion in Supplementary). After developing RhoFold+, we rigorously benchmarked and evaluated its performance across a broad range of tests (Fig.1b).
\begin{figure*}[!t]
    \centering
    \includegraphics[width=0.80\textwidth]{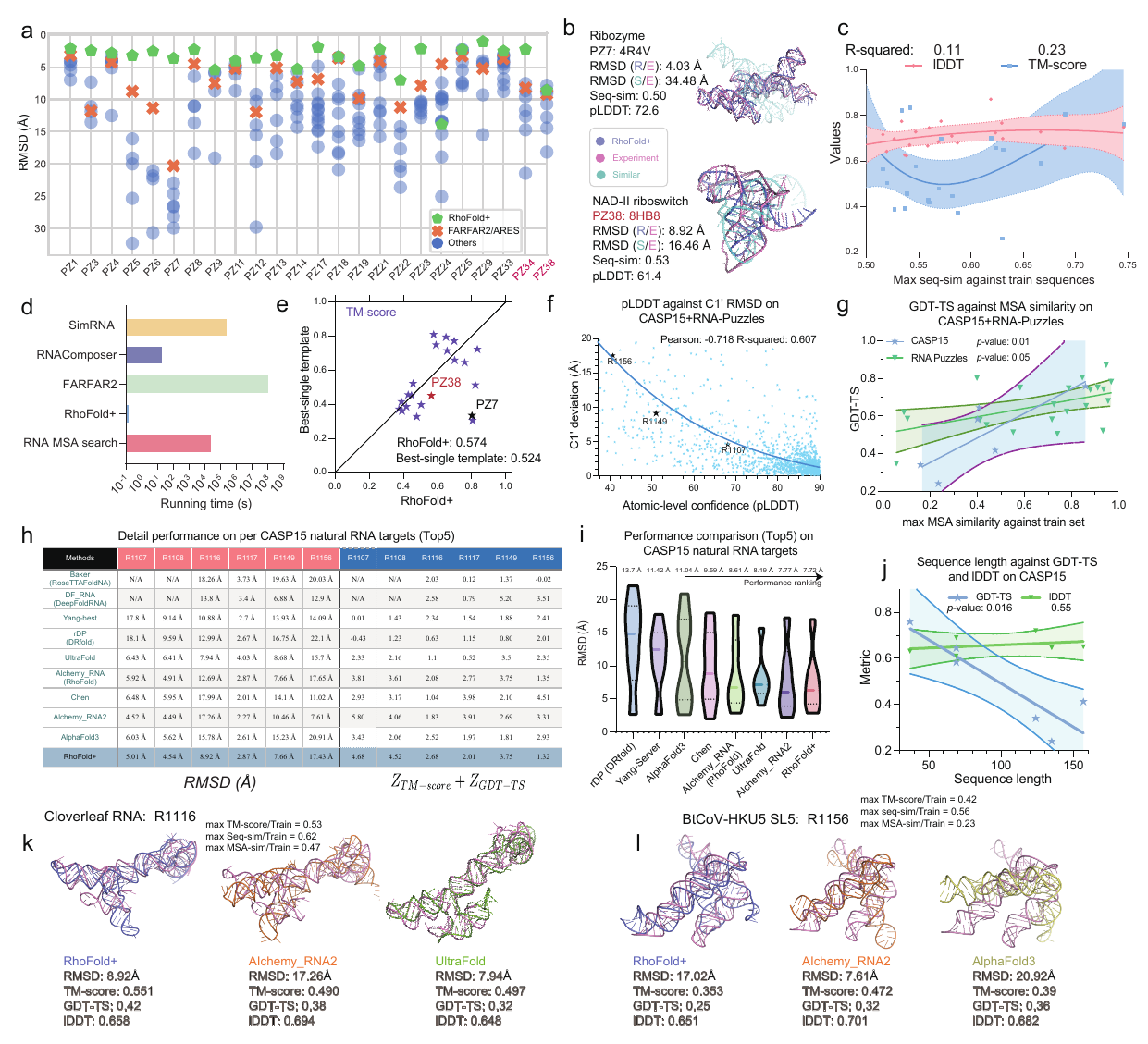}
    \caption{\textbf{Benchmarking RhoFold+ on previously held community-wide challenges}. The central curve in the \textbf{c}, \textbf{g}, and \textbf{j} panels represents the fitted regression model, while the two surrounding curves indicate the 95\% percentile intervals.
\textbf{a.} RMSD performance scatterplot of RhoFold+ and other methods across 24 non-overlapping, non-redundant RNA-Puzzles targets. Each point represents a predicted model from a specific method. 
    \textbf{b.} Visualization of RNA-Puzzles 7 and 38. In addition to the aligned RhoFold+ prediction, we show the most similar training structure with respect to each target, suggesting that RhoFold+ neither overfits the training set nor simply reproduces the most similar structure to the target. 
    \textbf{c.} Regression plot of the TM-score and lDDT of RhoFold+'s predictions against the maximum sequence similarity among all the training sequences, across all RNA-Puzzles targets. Each point represents an RNA-Puzzles target.
    \textbf{d.} Running time comparison for different methods. 
    \textbf{e.} Comparison of RhoFold+'s predictions against the respective best single templates from our training set across all RNA-Puzzles targets. 
    \textbf{f.} Regression plot for C1' RMSD against atom-level pLDDT across all RNA-Puzzles and CASP15 targets. 
    \textbf{g.} Regression plot for structure GDT-TS  against MSA similarity across all RNA-Puzzles and CASP15 targets.
    \textbf{h.} Detailed performance comparison for CASP15 natural RNA targets. The pink columns record detailed RMSD values and the blue columns record the sum of Z-scores for GDT-TS and TM-score.
    \textbf{i.} Comparison of RhoFold+'s average performance against the average reported performance of CASP15 groups and published works on CASP15 natural RNA targets.
    \textbf{j.} Regression plot for structure GDT-TS and lDDT against sequence length across all CASP15 targets.
    \textbf{k.} Comparison of RhoFold+'s predictions against AIchemy\_RNA2 and UltraFold on the R1116 target from CASP15. 
    \textbf{l.} For the R1156 target, showing one of RhoFold+'s potential failure cases involving incorrect stacking patterns and orientations.}
    \label{fig:overview}
\end{figure*}

\paragraph{Benchmarking RhoFold+ on RNA-Puzzles}
We performed a comprehensive retrospective comparison between RhoFold+ and other existing computational methods on two previously held community-wide challenges---RNA-Puzzles and CASP15. We first used the results from the RNA-Puzzles \cite{cruz2012rna,miao2015rna, miao2017rna,miao2020rna,magnus2020rna} competition, where the submissions were produced and optimized by human knowledge or computational methods.
Importantly, here RhoFold+ was trained using non-overlapping training data with respect to the RNA-Puzzles targets tested (see \emph{Methods}). We conducted preprocessing to obtain 24 single-chain RNA targets and excluded RNA complexes. This set of RNA targets contained two puzzles (PZ),  PZ34 and PZ38, that were introduced after our development of RhoFold+ (Fig.2a, Supplementary Figure 3) and thus served as a blind test. After collecting the predictions of other methods from the official server (http://www.rnapuzzles.org/), we found that the performance of RhoFold+ surpassed that of all other methods, including FARFAR2/ARES, on nearly all targets, except for PZ24. Notably, RhoFold+ outperformed the second-best method on more than half of the targets by $\sim$4 \AA{} RMSD. On 17 targets, RhoFold+ achieved RMSD values $<5$ \AA{}, and only one target exhibited an RMSD $>10$ \AA{} (Fig.2a, Supplementary Table 5). As a whole, RhoFold+ produced an average RMSD of 4.02 \AA{}, 2.30 \AA{} better than that of the second-best model (FARFAR2: top 1\%, 6.32 \AA{}). Assessed using the template modeling (TM)-score \cite{zhang2005tm}, RhoFold+ achieved an average of 0.57 (Supplementary Table 5), higher than the scores of other top performers (0.41 and 0.44). 

To show that the promising results on RNA-Puzzles did not arise from overfitting, we studied whether the sequence similarity between the test set and our training data was substantially positively correlated with RhoFold+'s performance, as measured by the TM-score and the local Distance Difference Test (lDDT; a superposition-free score that evaluates local distance differences for all atoms in a model \cite{zhang2004scoring, mariani2013lddt}). Such a correlation was previously found in protein structure prediction \cite{af2}; yet, here we found that $R^2$ values, which represent whether the slope is significantly non-zero, were 0.23 for the TM-score and 0.11 for the lDDT (Fig.2b,c), indicating no significant correlation between model performance and the similarity of our training and testing sets.
These results suggest that RhoFold+ can generalize in predicting accurate RNA structures. 
A case study of a representative RNA-Puzzles target, PZ7 (a 186-nucleotide-long Varkud satellite ribozyme RNA), exemplifies this finding. Here, the structure of the most similar RNA in the training set differed significantly from the structure of PZ7 (Fig.2b): the RMSD between these structures was 34.48 \AA{}. As another example, PZ38 exhibited the highest sequence similarity of 53\% with respect to all RNAs in our training set, and the RMSD between the structure of the most sequence-similar RNA and PZ38 was 16.46 \AA{} (Fig.2b). This was larger than the RMSD of 8.92 \AA{} between PZ38 and RhoFold+'s prediction.

In order to test RhoFold+'s ability to generalize for structure- (in addition to mainly sequence-) dissimilar targets, we sought to determine whether RhoFold+'s predictions could surpass the best single template (the most structurally similar model) in the training set for a given query. To investigate this, we compared the TM-scores between our predictions and experimentally determined structures against the TM-scores between the best single templates and experimentally determined structures across all RNA-Puzzles. For the majority of puzzles, RhoFold+ produced predictions with a higher global similarity and an average TM-score of 0.574, surpassing the best single template by 0.05 (Fig.2e, Supplementary Table 13). It is important to highlight that for proteins, surpassing the best single template required substantial progress. Indeed, it was only during CASP14 that computational methods outperformed the best single template. Although RhoFold+ generated considerably more accurate predictions than other methods under the conventional sequence similarity data splitting paradigm, we further tested RhoFold+'s adaptability by eliminating 3D structures from the training set whose TM-score, with respect to any target, surpassed a specified threshold (Supplementary Figure 6, Supplementary Table 6, 10). Even under this more demanding condition, RhoFold+ continued to exhibit promising performance (Supplementary Table 10).

In applying computational models to large-scale, real-world settings, speed is often a top priority. In addition to generating largely accurate folding results, we found that RhoFold+ is fast, with typical RNA-Puzzles predictions completed within $\sim$0.14 seconds (Fig.2d). In contrast, other approaches, including SimRNA \cite{boniecki2016simrna}, FARFAR2 \cite{watkins2020farfar2}, and RNAComposer \cite{biesiada2016automated}, exhibited significantly longer running times, likely due to the large-scale sampling processes employed by these methods (Fig.2d).

\paragraph{Benchmarking RhoFold+ on CASP15 targets}

As RNA-Puzzles was first released over a decade ago \cite{cruz2012rna}, we next used RhoFold+ to predict RNA targets from the more recent CASP15 \cite{casp, das2023assessment}. We focused on CASP15's six natural RNA targets (Fig.2h, Supplementary Figure 4). Artificially designed targets, which fell outside RhoFold+'s expected domain of application, were not included: in particular, the excluded targets were characterized by their lack of homology and divergence from our training set or their being RNA-protein complexes. We followed CASP15 guidelines, which specified that participating teams were permitted to submit up to five models. Utilizing different, randomly sampled MSAs (see \emph{Methods}), we modeled five candidate structures for each target using RhoFold+ and considered only the highest-performing prediction (Supplementary Table 6).

Several top-ranking CASP15 groups and recent published works on RNA 3D structure prediction \cite{baek2023accurate,li2023integrating,abramson2024accurate,wang2023trrosettarna,pearce2022novo} were included in our benchmarking. Particularly, CASP15 groups were divided into two categories, `server' and `expert', depending on whether or not human expert knowledge and fine-tuning were used. Regardless of category, many CASP15 groups employed computational pipelines that were based on comparative or statistical learning for natural targets, thus allowing us to assess RhoFold+'s learning capability. 
Our preliminary model, AIchemy\_RNA (RhoFold), was a participant in the `expert' category. Building on RhoFold, RhoFold+ represents a fully automated and end-to-end pipeline that is more similar to participants in the `server' category. 
Here, we found that RhoFold+ outperformed RhoFold on CASP15's natural RNA targets by an average RMSD of $\sim$1 \AA{}. 
Furthermore, RhoFold+ outperformed other methods whose predictions were available for all six natural RNA targets, including the first-ranked AIchemy\_RNA2, the second-ranked Chen method, and other computational methods, including DRfold \cite{li2023integrating}, DeepFoldRNA \cite{pearce2022novo}, AlphaFold3 \cite{abramson2024accurate} and trRosettaRNA \cite{wang2023trrosettarna} (Fig.2h,i).
Although RhoFold+ outperformed AIchemy\_RNA2 marginally by 0.06 \AA{} (average RMSD; Fig.2i), AIchemy\_RNA2 required expert knowledge. Additionally, RhoFold+ demonstrated accuracy comparable to each top-performing method on almost every natural RNA target, with the exception of R1156 (Fig.2h).

Following CASP15's assessment approach \cite{das2023assessment}, we also computed Z-scores for the predictions from all participating groups. CASP15 prioritized the TM-score and GDT-TS (Global Distance Test-Total Score, which evaluates both overall structure similarity and local alignment), leading us to assess these models based on the cumulative Z-scores of these metrics (Fig.2h). 
On the six natural RNA targets and among the subset of all CASP15 participants ranked on these specific targets, RhoFold (AIchemy\_RNA) was fourth, while RhoFold+'s performance was on par with AIchemy\_RNA2's (with a difference of 0.4 in the Z-score) and surpassed that of other methods. In a detailed analysis of performance on specific targets, we found that, for target R1108, RhoFold+ achieved the best Z-score and RMSD. 
Interestingly, RhoFold+ also attained the best Z-score for R1116, although its RMSD was $\sim$1 Å higher than that of UltraFold (other methods produced predictions with significantly lower accuracy, with RMSDs $>$10 Å). Upon further investigation, we found that, while UltraFold outperformed RhoFold+ on this metric by producing accurate local predictions, the predicted global structure was less accurate, as evidenced by a TM-score of 0.497 and a GDT-TS score $<$0.4. In contrast, RhoFold+ inaccurately predicted a helix angle, resulting in an RMSD of 8.92 Å, but its correctly predicted topology resulted in a higher TM-score of $>$0.55. For this target, AIchemy\_RNA2 incorrectly predicted the stem stackings and RNA topology, resulting in a high RMSD of 17.26 Å and a TM-score of $\sim$0.49. Notably, RhoFold+'s prediction for R1116 did not arise from overfitting, as indicated by the low maximum structural similarity (TM-score) and maximum sequence similarity (seq-sim) of R1116 with respect to the training set (Fig.2k, Supplementary Table 6).

We also looked into targets where RhoFold+ may achieve reduced performance and found that higher MSA quality correlated with better performance. While RhoFold+ accurately predicted local structural topologies, it struggled with aligning helices, particularly at junctions. This discrepancy may be due to the dynamic and flexible nature of RNA junctions, which often adopt multiple conformations \cite{gupta2022alternative,zhang2007visualizing,ding2023visualizing}, making them challenging for fully automated models to represent accurately (Fig.2k,l; detailed discussion in Supplementary).

\paragraph{Factors influencing prediction accuracy}

Building on the findings above, we performed a more comprehensive study involving all CASP15 natural RNAs and RNA-Puzzles targets. We observed that the prediction accuracy of RhoFold+ is sensitive toward the query's MSA profile similarity (see Supplementary \textit{Additional metrics}) against the training set (Fig.2g) and the complexity of RNA structures (query length; Fig.2j). Additionally, pLDDT scores were found to correlate with RhoFold+'s confidence, providing a useful metric for identifying regions with lower prediction accuracy, especially in more complex or less homologous queries (Fig.2f; detailed discussion and analysis in Supplementary).

\begin{figure*}[!t]
    \centering
    \includegraphics[width=0.70\textwidth]{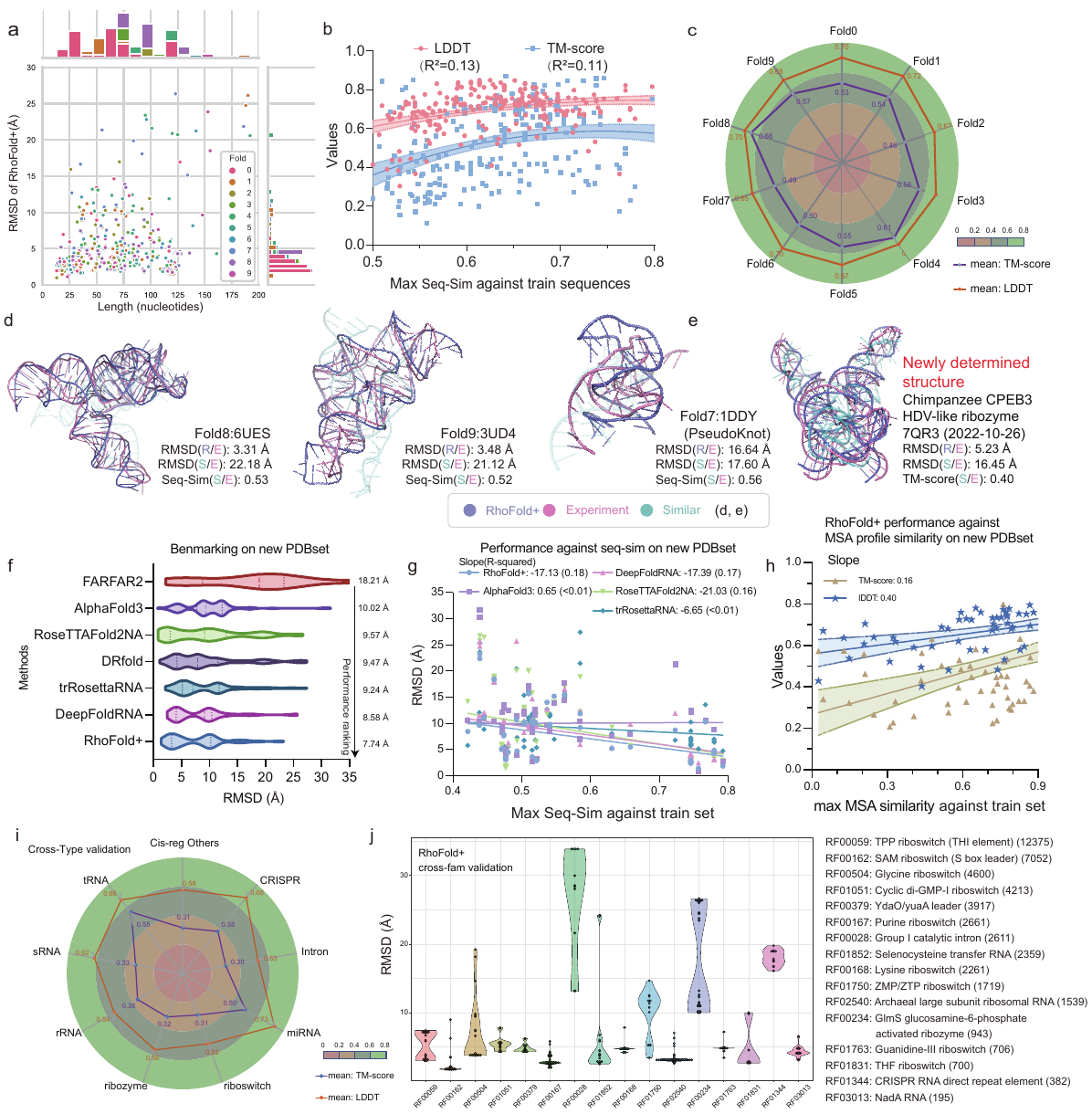}
    \caption{\textbf{Benchmarking RhoFold+ on all experimentally determined RNA structures supports RhoFold+'s accuracy and ability to generalize to unseen structures.} The central curve in the \textbf{b} and \textbf{h} panels represents the fitted regression model, while the two surrounding curves indicate the 95\% percentile intervals.
    \textbf{a-d}. Ten-fold cross-validation of RhoFold+ using all experimentally determined RNA structures. 
    \textbf{a.} Plot of RMSD values against sequence length for all cross-validation experiments. Each point represents an RNA structure and is colored according to the cross-validation fold. 
    \textbf{b.} Regression analysis for each prediction's TM-score (blue) and lDDT (pink) against the maximum sequence similarity with respect to all training data. Each point represents an RNA structure.  
    \textbf{c.} Average TM-score and lDDT for each fold. 
    \textbf{d.} Visualization of two representative riboswitch structures, 6UES and 3UD4, and a pseudoknot 1DDY (pink), along with the corresponding RhoFold+ predictions (slate) and the training RNA structures with the highest sequence similarity (cyan). 
    \textbf{e.} Visualization of a newly determined RNA structure, 7QR3, which has a low structural similarity with respect to the training set, but whose structure (pink) is accurately predicted by RhoFold+ (slate). The most similar structure, 7DLZ, is shown in cyan. 
    \textbf{f.} Comparison of average RSMD values generated by RhoFold+ and other methods on the new PDB set, a set of 76 newly determined solo RNA structures.
    \textbf{g.} Regression plot of the prediction RMSD values against maximum sequence similarity to the training set for RhoFold+ and other baseline methods.
    \textbf{h.} Regression plot of the correlation between RhoFold+ predictions' TM-score/lDDT and the maximum MSA profile similarity against the training set.
    \textbf{i.} Overview of cross-type validation performance of RhoFold+ measured by lDDT and TM-score. All structures in the type used for validation were masked during model training. 
    \textbf{j.} Violin plot of RhoFold+'s RMSD values in the cross-family validation. Here, all the structures in a family to be tested were masked during model training, and RhoFold+ accurately predicted RNA structures from most unseen families. The numbers of sequences in each family are shown in parentheses.}
    \label{fig:overview}
\end{figure*}

\paragraph{Benchmarking RhoFold+ on all determined RNA 3D structures}

After benchmarking RhoFold+'s with RNA-Puzzles and CASP15, we next evaluated RhoFold+ in greater detail using all experimentally determined RNA structures, as defined by the BGSU Representative Sets of RNA structures (preprocessed to remove redundancy). To further study RhoFold+'s performance, we performed ten-fold cross-validation by iteratively masking 80 sequence clusters for validation and leaving 702 sequence clusters for training. We found that RhoFold+'s performance across all RNA structures was robust regardless of the train-test data split and fairly consistent across all folds (Fig.3a,b,c). Slight variations in TM-score might be caused by challenging targets such pseudoknot cases in Fold2 and Fold7 similar to PZ24 (Fig.3c,e), and we expect that RhoFold+'s predictions on such targets could be improved if secondary structure constraints were provided. Also, during our cross-validation test, RhoFold+'s accurate predictions were not due to merely mimicking the most sequence-similar training data (Fig.3b,d,e). A plot of the RMSD against the sequence length shows that RMSD values were largely distributed below 10 \AA{}, independent of the sequence length (Fig.3a). Outliers with RMSD $>20$~\AA{} were more likely to occur for sequences longer than 200 nt, where we expect further improvement by more tuning on long RNAs (detailed discussion in Supplementary).

As a further evaluation of RhoFold+'s capabilities, we considered the model's performance on newly determined RNA single-stranded structures released subsequent to the compilation of our training dataset. This approach acted as an additional blind test, similar to the CASP15 competition. We included comparisons against FARFAR2 and recent deep learning methods~\cite{li2023integrating, abramson2024accurate, wang2023trrosettarna, pearce2022novo}, all of which have inference code and/or servers available and some of which also participated in CASP15 (see \textit{Methods}).  RhoFold+ outperformed all benchmarked models, achieving the highest average accuracy as measured by RMSD. RhoFold+ produced an average RMSD of 7.74 \AA{}, which was approximately 0.8 \AA{} and 10.5 \AA{} better than the second-ranked DeepRNAFold and the lowest-ranked FARFAR2, respectively. Notably, on average, RhoFold+ also outperformed AlphaFold3 and RoseTTAFold2NA by approximately 2.2 \AA{} and 1.8 \AA{}, respectively (Fig.3f; detailed discussion in Supplementary). 
 These results were consistent with the performance observed in our previous benchmark on CASP15, suggesting that RhoFold+ accurately generalizes to newly determined structures not seen in our training set. Furthermore, these results support that AlphaFold3 and RoseTTAFold2NA, which are designed to predict biomolecular complexes, do not perform as well as RhoFold+ when applied to single RNA molecules. Further examining sequence and structural similarities to our training set reveals that RhoFold+ maintained strong performance even with sequence similarities below 0.5 (Fig.3g), and TM-score was much influenced by  MSA profile similarity while local accuracy (lDDT) remained high and robust (Fig.3h). Additionally, RhoFold+ demonstrated strong generalizability, accurately folding structures like 7QR3 despite its low similarity to the closest training template, 7DLZ (Fig.3e, TM-score: 0.40, RMSD: 16.45 Å).

\paragraph{RhoFold+ generalizes to unseen RNA types and families}
Having demonstrated that RhoFold+ can generalize to predicting RNA structures with divergent sequence similarities, structural similarities, and dates of release, we next investigated RhoFold+'s ability to handle different RNA types and families defined by expert knowledge. In particular, RNA types and families---such as those curated in Rfam \cite{griffiths2003rfam}---are often classified manually based on factors including function, structure, and co-evolutionary information. Addressing the challenge of generalizing to different RNA types and families may be considerably more demanding for deep learning methods like RhoFold+, as such a task requires larger domain shifts.

We benchmarked RhoFold+'s cross-type performance by training the model on a subset of all RNA types while testing on the others. RhoFold+ showed robustness across RNA types. Though struggling with introns and riboswitches, it performed well on tRNA and miRNA types, achieving TM-scores up to 0.73 (Fig.3i). When compared to FARFAR2, RhoFold+ outperformed it across all RNA types, particularly in tRNAs and rRNAs, with smaller margins for riboswitches (detailed discussion in Supplementary). For cross-family tests, RhoFold+ achieved an average RMSD of 6.69 Å (Fig.3j), but struggled with complex families like group I introns (RF00028). This difficulty is consistent with challenges observed in cross-type tests, such as for complex RNA types like introns and CRISPR RNA elements (RF01344). These elements interact with various proteins and enzymes, and focusing solely on RNA structure without considering these interactions may limit the prediction accuracy (detailed discussion in Supplementary). Overall, these tests demonstrate RhoFold+'s ability to generalize across unseen RNA types and families, though challenges remain for complex structures and datasets with limited available data.

\begin{figure*}[!t]
    \centering
    \includegraphics[width=0.7011\textwidth]{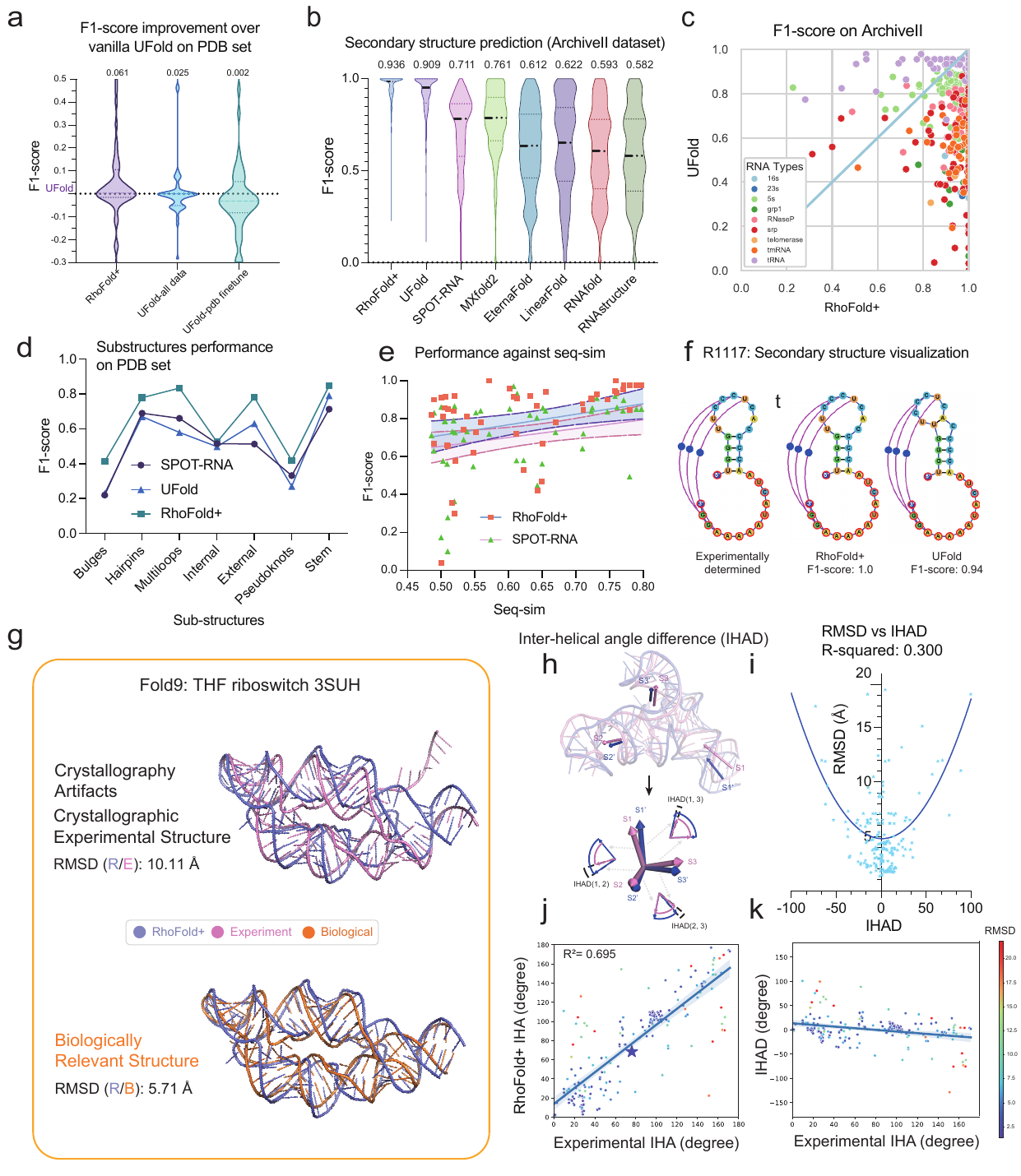}
    \caption{\textbf{RhoFold+ accurately predicts secondary structures and inter-helical angles from experimental data.} The central curve in the \textbf{e}, \textbf{j}, and \textbf{k} panels represents the fitted regression model, while the two surrounding curves indicate the 95\% percentile intervals.
    \textbf{a.} F1-score comparison against multiple configurations of UFold on the PDB set. Here, a version of UFold trained on bpRNA is also presented as a baseline, in order to evaluate the improvement in terms of F1-score. 
    \textbf{b.} F1-score distribution of various methods on the ArchiveII dataset. Average scores are indicated at the top of the plot. 
    \textbf{c.} F1-score comparison between RhoFold+ and UFold on the ArchiveII dataset. Each point represents an RNA structure and is colored according to its RNA type. 
    \textbf{d.} F1-score comparison of RhoFold+ vs. UFold and SPOT-RNA on RNA substructures in the new PDB set.
    \textbf{e.} F1-score comparison of RhoFold+ vs. UFold and SPOT-RNA against sequence similarity of RNA structures in the new PDB set.
    \textbf{f.} Visualization of a CASP15 RNA target where RhoFold+ predicted the correct secondary structures including pseudoknots.
    \textbf{g.} Visualization of a swapped dimer, 3SUH, for which RhoFold+'s prediction (purple) resembles the biologically meaningful structure (orange) instead of the crystallographic artifact found in the PDB (pink). 
    \textbf{h.} Visualization showing the definition of the inter-helical angle difference (IHAD), which is the difference between the inter-helical angles (IHAs) derived from RhoFold+'s prediction and the experimentally determined structure. 
    \textbf{i.} Regression analysis between the IHAD and RMSD of RhoFold+'s predictions. Each point represents an RNA. 
    \textbf{j.} Comparison between the IHAs derived from RhoFold+'s predictions against those from experimental structures. Each point represents an angle instance and is colored according to the RMSD between the experimental structure containing the angle and the structure predicted by RhoFold+. 
    \textbf{k.} Plot of the IHAD against experimentally determined IHA values. The coloring is the same as in \textbf{j}.}
    \label{fig:overview}
\end{figure*}

\paragraph{RhoFold+ predicts secondary structures and substructures}

RhoFold+ can accurately predict RNA 3D structures, but the limited number of experimentally determined RNA structures and types makes it difficult to understand the space of all possible RNA folds. This is particularly true for complicated and large RNA types, including internal ribosomal entry sites (IRESs), introns, synthetic RNAs, and long non-coding RNAs (lncRNAs). RNA secondary structures, however, can be more easily determined in experiments, and accurate secondary structure predictions can supplement the predictions of 3D structures, offering valuable insights into RNA folding and function. Therefore, we adapted RhoFold+ to predict secondary structures as well.
As RhoFold+ was designed to predict RNA 3D structures, we incorporated a post-processing module that utilizes the features retrieved from RhoFold+'s Rhoformer to predict secondary structures (since Rhoformer's features show attention maps highly aligned with the contact maps; Supplementary Figure 8 and Supplementary Table 14).
This module takes into account the same structural information as the module performing 3D reconstruction but operates under distinct geometric and biological constraints imposed to predict secondary structure.

We benchmarked RhoFold+'s performance on newly determined PDB structures (the `new PDB set') and the ArchiveII dataset \cite{fu2022ufold}, which includes secondary structure information for diverse RNAs. On the new PDB set, RhoFold+ outperformed UFold~\citep{fu2022ufold} by 0.035 in the average F1-score (Fig.4a), even when UFold was trained on all available data (bpRNA-1M and PDB). On the ArchiveII dataset comprising 2,975 RNA samples, RhoFold+ also outperformed other secondary structure prediction methods (Fig.4b), particularly on larger RNA types (Fig.4c). For instance, it achieved an F1-score of 0.60 on structured domains in the dengue virus transcriptome (Supplementary Table 19), aligning with results from mutational profiling (RING-MaP) \cite{dethoff2018pervasive,rice2014rna}. Similarly, RhoFold+’s strong performance did not stem from mimicking training data, as it maintained an F1-score of $\sim$0.7 even when sequence similarity dropped below 50\% (Fig.4e), and achieved a perfect F1-score of 1.0 on the CASP15 target R1117 (Fig.4f). These results suggest that RhoFold+ not only excels in predicting 3D structures but also generates rich, meaningful representations that enable state-of-the-art secondary structure prediction.

We further evaluated substructures within RNA secondary structures, finding that RhoFold+ consistently outperformed SPOT-RNA~\citep{singh2019rna} and UFold~\cite{fu2022ufold} across all substructures, with the most significant improvements in multiloops and external loops, while internal loops and pseudoknots showed similar performance across methods (Fig.4d). These results underscore RhoFold+'s potential capability in predicting RNA secondary structures and enhancing our understanding of RNA function.

\begin{figure*}[!t]
    \centering
    \includegraphics[width=0.85\textwidth]{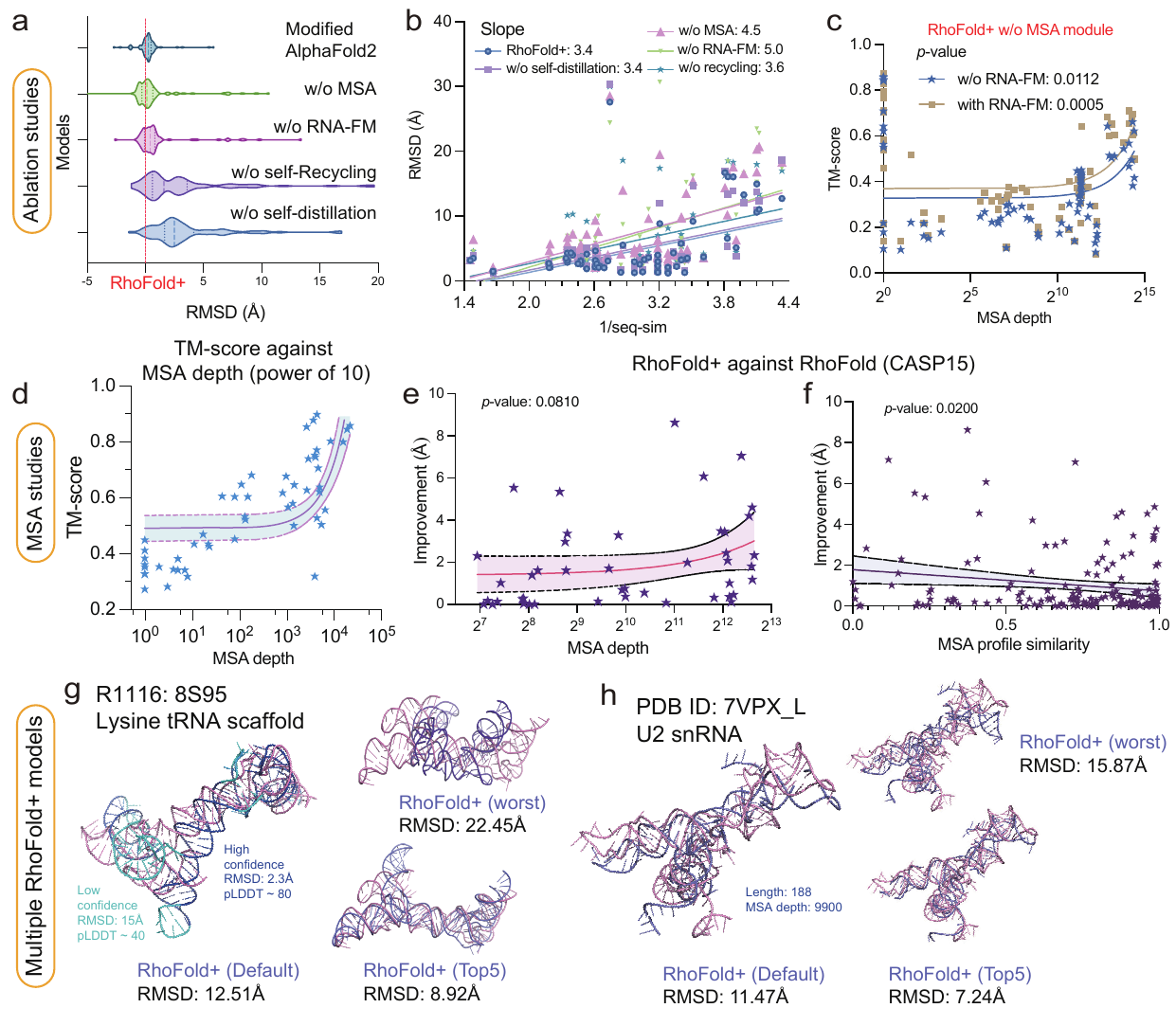}
    \caption{\textbf{Ablation studies of RhoFold+ and sampling of multiple models.} The central curve in the \textbf{e} and \textbf{f} panels represents the fitted regression model, while the two surrounding curves indicate the 95\% percentile intervals.
    \textbf{a-c}. Ablation studies of RhoFold+ using the Ablation set (excluding complexes).
    \textbf{a.} Ablation studies of RhoFold+ after removing corresponding modules in RhoFold+ with performance measured by RMSD.
    \textbf{b.} Regression analysis for prediction accuracy (measured by RMSD) against the reciprocal of sequence similarity. 
    \textbf{c.} A regression analysis of the TM-score against MSA depth for the ablation study of the RNA-FM module. Note that the \textit{x}-axis is log-scaled. 
    \textbf{d-f}. Detailed studies on the impact of MSAs.
    \textbf{d.} Plot of prediction accuracy (measured by the TM-score) against MSA depth.
    \textbf{e.} Plot of the improvement of RhoFold+ against RhoFold (measured by RMSD) across different MSA depths.
    \textbf{f.} Plot of the improvement of RhoFold+ against RhoFold (measured by RMSD) across different MSA profile similarities.
    \textbf{g-h.} Examples of RhoFold+'s multiple models.
    \textbf{g.} Visualization of a CASP15 target where RhoFold+ produces an RMSD of 12.51 \AA{}, but improves by 8.92\AA{} using Top5 prediction from MSA sampling. 
    \textbf{h.} Visualization of a newly determined RNA structure where RhoFold+'s RMSD improves by 7.92\AA{} using Top5 prediction from MSA sampling.}
    \label{fig:overview}
\end{figure*}

\paragraph{Correcting artifacts and inter-helical angle prediction}
As RhoFold+ accurately predicts RNA structures at both the secondary and tertiary levels, we asked whether we could leverage RhoFold+ for experimental efforts. Toward this, we investigated two use cases of RhoFold+: (1) for correcting experimental structural artifacts and (2) for guiding RNA construct engineering. 

X-ray crystallography is widely used to resolve RNA 3D structures, but it can introduce artifacts like domain-swapped dimers \citep{bou2020structural}, potentially misleading machine learning models that do not generalize well. In one case, RhoFold+'s prediction for 3SUH initially yielded a high RMSD of 10.11 Å compared to the PDB structure. However, further analysis revealed that the crystal structure involved a domain-swapped dimer. When comparing RhoFold+'s prediction to the inferred monomeric structure, the RMSD improved to 5.71 Å, indicating RhoFold+ accurately predicted the biologically relevant structure (Fig.4g). Similar findings were also observed for the ZTP riboswitch \cite{trausch2015metal} (Supplementary Figure 9), suggesting RhoFold+ can effectively correct for such experimental artifacts.

When comparing experimental data with RNA 3D models, additional geometric metrics, like inter-helical angles (IHAs), can provide insights beyond standard global alignment measures such as RMSD, lDDT, and TM-score. IHAs, which can be estimated using experimental methods, are useful for validating predicted models and guiding RNA nanostructure design. We introduced the IHA difference (IHAD) as a metric to benchmark RhoFold+'s predictions (Fig.4h; see Supplementary \textit{Additional metrics}), finding that IHAD can reveal discrepancies in stem orientations that are not captured by RMSD alone (Fig.4i). Our analysis shows that RhoFold+ generally predicted stem directions accurately (Fig.4j,k), though performance decreased for IHAs near 0° or 180°, probably due to underfitting of parallel stems in large and complex structures (Fig.4k; detailed discussion in Supplementary). We further demonstrated the practical application of IHAs by predicting values for RNA constructs like the FMN riboswitch and the P4-P6 domain from the Tetrahymena group I intron (Supplementary Figure 9).

\paragraph{Ablation studies and generation of multiple predictions} 
Given RhoFold+'s high accuracy and speed, we finally conducted ablation studies to understand which components and information are important to RhoFold+’s predictions. The architectural components we investigated included 4 different modules (Fig.5a, see \textit{Methods}). Ablation studies were performed on 138 PDB targets (collected between April 2022 and December 2023) with sequence similarities below 80\% to our training set and lengths ranging from 16 to 300 nt (the `Ablation set'). By removing each RhoFold+ component, we observed that all contributed to improving the performance, with the MSA module being the most critical, followed by the RNA-FM language model (Fig.5a). The RNA-modified version of AlphaFold2, without the MSA module, performed worse than RhoFold+ (Fig.5a). Notably, removing RNA-FM led to a sharper performance decline for dissimilar sequences (Fig.5b), and the RNA-FM module seemed to compensate for the loss of the MSA module, maintaining higher TM-scores (Fig.5c). Additionally, removing the recycling module most significantly affected predictions for longer sequences, likely due to its role in effectively deepening the model (Supplementary Figure 7; detailed discussion in Supplementary).

These findings are consistent with our results for CASP15's natural RNA targets and RNA-Puzzles, where MSA quality significantly impacts predictions. We also explored how the number of sequences in the extracted MSA influences accuracy. While RhoFold+ is limited to 256 MSAs due to training constraints, this limit did not compromise its effectiveness. A key enhancement in RhoFold+ is its ability to generate multiple predictions by sampling or clustering from a fixed number of MSAs, allowing for broader prediction selection and improved outcomes. Performance on RNA-Puzzles showed an inverse correlation with reduced MSA counts, with a marked improvement when the MSA number exceeded 100 (Fig.5d), indicating that a larger MSA pool enhances model optimization (detailed discussion in Supplementary).
With this expanded MSA sampling, the lowest RMSD of RhoFold+'s Top5 predictions significantly decreased compared to RhoFold's, correlating positively with increased MSA depth and yielding an up to 10 Å improvement (Fig.5e). This improvement was more pronounced when the MSA profile similarity between the query and training sequences was high, resulting in smaller gains when similarity was already strong (Fig.5f). Overall, additional MSA sampling is crucial for high performance, as demonstrated for CASP15 target R1116 and PDB 7VPX\_L (Fig.5g,h).

%% file: discussion.tex
In this work, we have developed an end-to-end language model-based deep learning method, RhoFold+, to predict RNA three-dimensional structure from sequence. RhoFold+ is a fully automated and differentiable model that integrates an RNA language model pre-trained on $\sim$23.7 million RNA sequences without structural information leakage and multiple strategies to augment the scarce training data. RhoFold+ outperforms other RNA structure prediction approaches based on deep learning on CASP15 natural RNA targets and achieves a sub-4 \AA{} mean RMSD for the non-overlapping and non-redundant RNA-Puzzles structures. Because RhoFold+ does not require any time-consuming and computationally intensive sampling processes, RhoFold+ is also fast and efficient; neither does it rely on expert knowledge, which has been used in the most high-performing approaches to RNA structure prediction to date. RhoFold+ is able to generalize from different sets of training data and accurately predict both available RNA 3D structures and newly determined ones, an observation underscored by RhoFold+'s strong robustness during cross-fold validation. Additionally, RhoFold+ can accurately predict unseen RNA structures during cross-family and cross-type validations. Although RhoFold+ was designed to predict 3D structures, it can also accurately predict RNA secondary structures. Applying RhoFold+ for the prediction of inter-helical angles---a task inspired by cryo-EM- and NMR-based construct engineering design---suggests its potential to accelerate the process of experimentally determining more RNA structures.

Although RhoFold+ shows promising performance, it shares limitations with other deep learning methods for RNA structure prediction. First, our knowledge of RNA structural diversity is limited, making it challenging to predict different conformations of the same RNA molecule due to their dynamic nature and interactions with other molecules. RNA junctions, for example, can adopt multiple conformations and are better represented as dynamic ensembles \cite{gupta2022alternative,zhang2007visualizing,ding2023visualizing}. Second, due to insufficient data, predicting large and complex RNA structures, particularly those with multiple helices or pseudoknots, remains difficult, especially for sequences longer than ~500 nucleotides. Third, RNA complexes involving ligands or proteins present additional challenges, as current methods often fail to account adequately for these interactions, reducing accuracy. While methods like AlphaFold3 \cite{abramson2024accurate} and RoseTTAFoldNA \cite{baek2023accurate} can predict RNA complexes, their accuracy is still limited, and they perform less well than RhoFold+ on single-strand RNAs. 
\highlight{Fourth, RhoFold+ and similar models are trained on datasets derived from specific environmental conditions, which may not generalize well to the diverse and dynamic solution conditions that RNA molecules encounter \textit{in vivo}. These conditions include varying concentrations of ions, such as magnesium and potassium, and the presence of ligands, which are known to play critical roles in RNA folding and stability.}

Methods that rely on MSAs are limited by the availability of these alignments, making accurate predictions difficult for artificially designed or orphan RNAs lacking corresponding MSAs. Although RNA-FM has helped mitigate this dependency, challenges remain. RhoFold+ and similar deep learning models, while accurate, are hindered by limited knowledge of RNA structural diversity, difficulties in predicting large and complex structures, and the reliance on MSAs. To mitigate these obstacles, integrating probing methods to define secondary structures, incorporating molecular dynamics and energy function techniques, and improving the MSA extraction process could potentially enhance RhoFold+'s accuracy. Additionally, addressing RNA-protein and RNA-ligand interactions remains crucial, and integrating RhoFold+ with protein structure prediction tools like RoseTTAFoldNA or AlphaFold3 could improve its capabilities in these areas.

%% file: methods.tex
\subsection*{The RhoFold+ platform}
\subsubsection*{Multiple sequence alignment feature generation} 
We used the MSAs constructed by Infernal \cite{nawrocki2013infernal} and rMSA (https://github.com/pylelab/rMSA) to capture co-evolutionary information of the sequence as an additional input. Using Infernal, it is possible to locate homologous sequences with conserved secondary structures; on the other hand, rMSA employs an iterative search strategy based on RNA sequence databases. We utilized the nucleic acid sequence databases Rfam and RNAcentral \cite{RNAcentral}. In AlphaFold2, a similar approach was used with different alignment tools and sequence databases. Given the need to produce several models and the constraints imposed by hardware memory, we reduced our fully extracted MSAs to a maximum of 256 sequences during the training phase. Subsequently, during the inference phase, 256 MSAs were either randomly selected or chosen through clustering, and then fed into RhoFold+. We implemented clustering with conserved secondary structure or sequence embeddings from our pre-trained RNA language model. Different sampled and clustered results can be thus used for multiple predictions, as marked by Top5, Top10, etc. By default, the top 256 MSAs are chosen as input features for predicting the standard structure, which we refer to as standard RhoFold+. RhoFold+ (TopK) refers to the optimal model selected from K different models generated using distinct sampled MSAs.

\subsubsection*{RNA-FM language model}
\paragraph{Overview of RNA-FM}
Our foundation model provides meaningful representations that are inferred from standalone sequence information. These representations may improve performances in various downstream tasks, especially for those with insufficient annotated data. Inspired by recent studies \cite{vaswani2017attention, rives2021biological}, we leverage a general transformer architecture. In particular, our framework was built on the bidirectional transformer language model proposed in BERT \cite{kenton2019bert}, followed by the unsupervised training scheme. We named our framework `RNA-FM' as it represents a foundational model for future RNA-related studies (Supplementary Figure 2).
Below, we detail how we constructed the large-scale non-coding RNA (ncRNA) dataset, followed by model and training details. 

\paragraph{Large-scale pre-training dataset} 
The large-scale dataset used in the pre-training phase was collected from RNAcentral \cite{RNAcentral}, the largest ncRNA dataset available to date. This dataset is a comprehensive collection of ncRNA sequences, representing all ncRNA types from a broad range of organisms. It combines ncRNA sequences across 47 different databases, resulting in a total of $\sim$27 million RNA sequences (Supplementary Table 2, 3, 4).

We preprocessed all ncRNA sequences by replacing `T's with `U's since they are both complementary to adenine and similar in structure (`T's for representing thymine in DNA, while `U's for uracil in RNA). This resulted in a dataset involving 4 main types of bases (16 counted types of combinations in total: `A', `C', `G', `U', `R', `Y', `K', `M', `S', `W', `B', `D', `H', `V', `N', and `-').
Moreover, to minimize redundancy without compromising the size of our dataset (i.e., to preserve as many sequences as possible), we removed duplicate sequences using CD-HIT-EST, which was set to a 100\% similarity threshold. After the above preprocessing steps, a final, large-scale dataset consisting of over 23.7 million ncRNA sequences was obtained. We named this final dataset `RNAcentral100', and we used this dataset to train our RNA foundation model in a self-supervised manner (see \textit{RNA language model (RNA-FM)} in the Supplementary Information for more details).

\paragraph{RNA-FM training details} 
Our RNA-FM framework comprises 12 transformer-encoder blocks, inspired by BERT \cite{kenton2019bert, vaswani2017attention}. Each block includes a 640-hidden-size feed-forward layer and a multi-head self-attention layer with 20 heads, along with layer normalization and residual connections applied pre- and post-block, respectively. For an RNA sequence of length $L$, RNA-FM takes raw sequential tokens as input, mapping each nucleotide into a 640-dimensional vector via an embedding layer, forming an $L \times 640$ embedding matrix. This matrix passes through each encoder block, retaining its size throughout, and is followed by a softmax layer to predict corresponding tokens, including 16 nucleotides and four specific functional identifiers. Additional model details are in the Supplementary Information.

During pre-training, we employed self-supervised training akin to BERT \cite{kenton2019bert}, randomly replacing 15\% of nucleotide tokens with a special mask token. If the $i$-th token was chosen, it was replaced with (1) the [MASK] token 80\% of the time, (2) a random token 10\% of the time, and (3) left unchanged 10\% of the time. We trained the model using masked language modeling (MLM) \cite{kenton2019bert}, predicting the original masked token via cross-entropy loss. This training strategy is formulated as an objective function as follows:

\begin{equation}
\mathcal{L}_{MLM}= \mathbb{E}_{x \sim \mathcal{X}}\mathbb{E}_{x_{\mathcal{M}}\sim x}\sum_{i \in \mathcal{M}} - \log p(x_i|x_{/\mathcal{M}}).  \label{EQ.1}
\end{equation}

A set of indices $\mathcal{M}$ is randomly sampled from each input sequence $x$, covering 15\% of the sequence, and the corresponding tokens are replaced with mask tokens. For each masked token, given the masked sequence ($x_{/\mathcal{M}}$) as context, the objective function minimizes the negative log-likelihood of the true nucleotide $x_i$. This approach captures dependencies between the masked and unmasked parts of the sequence, leading to accurate predictions for masked positions. Training with the objective function in Eq.~\ref{EQ.1} allows RNA-FM to effectively model representations of each sequential token. We trained RNA-FM on eight A100 GPUs (80 GB each) for one month, using an inverse-square-root learning rate schedule with a 0.0001 base rate, 0.01 weight decay, and 10,000 warm-up steps. To optimize memory usage and batch size, we set the maximum input sequence length to 1,024, accelerating the training process.

\subsubsection*{Efficient development of a self-distillation dataset}
Although our RNA-FM can alleviate the problem of data scarcity, there is still less structural data available for RNAs than for proteins. As a result, we collected a non-redundant, self-distillation dataset with ground-truth secondary structure from the RNAStralign and bpRNA-1M databases. We filtered this dataset by removing sequences with more than 256 or fewer than 16 nucleotides, resulting in a dataset of 27,732 sequences. RhoFold+ was initially trained using only PDB data, which was then used to generate a self-distillation dataset by inferring pseudo-structural labels. We re-trained the model by sampling 25\% of the PDB data and 75\% of the distillation data for further improvement. During training, we masked out pseudo-label residues with pLDDT scores \textless 0.7 and uniformly sub-sampled the MSAs to augment the distillation dataset.

\subsubsection*{A structure prediction module}
\label{sec:structure module}
RhoFold+'s structure module aims to predict the 3D structure of an RNA based on the sequence and pair representation extracted by Rhoformer. The structure module of AlphaFold2 directly predicts the rotation and translation matrices of the backbone frames, as these are the most influential factors in protein folding. However, RNA folding is primarily driven by nucleotide base pairing. Due to their irregular structural patterns, directly predicting the base frame ($C1^{'}, N1/N9, C2/C4$) defined over the nucleotides may pose a convergence problem in our experiments (Supplementary Table 1). To efficiently reconstruct the RNA full-atom coordinates, we used frame ($C4^{'}, C1^{'}, N1/N9$)
and four torsion angles $\alpha, \beta, \gamma, \omega$ to resolve this issue.
Supplementary Table 1 provides the definitions of torsion angles and corresponding rigid groups. The 3D positions are modeled using invariant point attention (IPA), a geometry-aware attention operation. Based on Rhoformer's output features and pair presentation, the IPA operation predicts the rotation and translation matrices for each frame. In addition, the predicted structure is refined iteratively using a recycling strategy, in which the Rhoformer receives the prediction from the previous iteration. The recycling process ends when the predicted lDDT (pLDDT), which is one of the outputs generated by IPA that measures the quality of the predicted 3D structure, converges. With the reconstructed full-atom coordinates, biological constraints, such as base-pairing, can be enforced directly in 3D space to optimize the structure module and generate biologically valid structural predictions.

\subsubsection*{Feature processing with Rhoformer}
As with the Evoformer introduced in AlphaFold2, our main module, Rhoformer, is composed of a series of transformer modules with gated self-attention layers, which are employed to learn evolutionary information and simultaneously update the pairwise sequence embeddings and MSA representations. A transition block comprising two linear layers is added to the resulting pair and MSA representations to increase the embedding dimension by a factor of four, thereby increasing the model's capacity. Lastly, four self-attention blocks are stacked on the Rhoformer to refine the pair and MSA representations. These representations are then fed into the structure module to obtain the predicted full atom coordinates in three-dimensional space, as described in the following sections.

\subsubsection*{The structure prediction loss}
The loss function is defined at 1D, 2D, and 3D levels. Each of these levels is discussed in detail below. We first employed a masked language modeling loss $L_{mlm}$ to improve the extraction of co-evolutionary information from the MSAs at the 1D level without adding curated correlation features. In our experiments, 5\% of the nucleotides were randomly masked, and a linear projection layer was utilized to reconstruct them.

At the 2D level, a distance loss $L_{dis}$ and a secondary structure loss $L_{ss}$ were applied to supervise RhoFold+ to learn the pairwise positional correlations between each residue. In particular, three feed-forward layers were used for distance prediction to predict the pairwise distance between the $P$, $C4$, and $N$ atoms. The distance was divided into 40 bins, where the first and last bins indicate $<2$~\AA{} and $>38$~\AA{}, respectively, and the distances between $[2$~\AA{}, 38~\AA{}$]$ were evenly divided into 36 bins. Additionally, the cross-entropy loss was used to determine if the distance predictions belong to the correct bin. For the secondary structure prediction loss $L_{ss}$, a feed-forward layer was leveraged on top of pairwise features to predict the secondary structure. The secondary structure $C$ is a $L\times L$ binary matrix, where $L$ denotes the sequence length, and $C_{i,j}=0$ or $1$ indicates if the $i$-th and $j$-th residue form a base pair.

At the 3D level, gradients were derived from the main Frame Aligned Point Error (FAPE) loss, the secondary structure constraint loss, and the clash violation loss $L_{clash}$. AlphaFold2's FAPE loss compares a set of predicted atom coordinates under a set of predicted local frames to the corresponding ground-truth atom coordinates and ground-truth local frames. Loss is independent of rigid motions. The loss remains constant when the predicted structure differs from the actual structure by arbitrary rotation and translation.

The secondary structure constraint loss, $L_{ss3d}$, encodes secondary structural information directly into 3D prediction. In order to unify the calculation of different types of base pairing constraints in 3D space, we introduced four fixed pseudo-atoms (T1, T2, T3, T4) in the local coordinate system of a base \cite{xiong2021pairing} (Supplementary Figure 1). $L_{ss3d}$ aims to constrain the pseudo atoms in two base-paired nucleobases to satisfy the base-pairing property (base-base interactions). For two residue $m$ and $n$, we computed the pairwise distance of the fixed points: $\mathbb{D}^{m,n}=\{d_{i,j}^{m,n}| i,j \in \{1,2,3,4\}\}$, where $m$, $n$ denote two RNA residues and $i$, $j$ are the indexes of the four atoms. We defined $L_{ss3d}$ as follows:

\begin{align}
    L_{ss3d} = \sum_{\genfrac{}{}{0pt}{}{m=1}{n=1}}^{N_{\text {nbpairs }}} \max \left(\hat{d}_{i,j}^{m,n}-\tau-\mathbf{d}_{i,j}^{m,n}, 0\right),
\end{align}
where $m$, $n$ are the indices of two residues that form a base pair, $i,j \in \{1,2,3,4\}$ denote the index of four pseudo atoms, $\hat{d}_{i,j}^{m,n}$ is the distance between two pseudo atoms $i$ and $j$ in the predicted structure,  $\mathbf{d}_{i,j}^{m,n}$  is the corresponding standard pairwise distance, $N_{\text {nbpairs }}$ is the number of all base pair residues in this structure, and $\tau$ is a tolerance distance threshold. The $L_{ss3d}$ penalizes pairwise atom distances in the nucleotides when two residues form a base pair. The calculation of the standard pairwise distance $\mathbf{d}_{i,j}^{m,n}$ is divided into two scenarios: 1) when the training sample comes from PDB data with 3D native structures, $\mathbf{d}_{i,j}^{m,n}$ comes directly from the structure; 2) when the training sample comes from self-distilled data, $\mathbf{d}_{i,j}^{m,n}$ are the statistical values generated from all PDB structures of the corresponding type of base pair. This can prevent RhoFold+ from overfitting the pseudo-labels and make full use of secondary structure information. 

$L_{clash}$ expects the model to learn to avoid atom clashes by penalizing distances that are too short between atoms according to their Van der Waals radii. Additionally, we employ a loss, $L_{pLDDT}$, to train an lDDT evaluator that scores the predicted 3D RNA models as an indicator for global recycling (as introduced above). The purpose of the $L_{pLDDT}$ loss is to train an lDDT evaluator that predicts the lDDT of the predicted 3D model based on the ground truth structure. The lDDT value is discretized with a 0.02-bin interval into 50 bins. Once a predicted 3D model has been generated, its lDDT is computed against the ground truth structure as the ground truth pLDDT label, and the lDDT evaluator generates the predicted pLDDT bin. Cross-entropy loss is used as $L_{pLDDT}$ to determine whether the predicted lDDT falls within the ground truth bin.

The overall loss function is:
\begin{align} 
\resizebox{.9\hsize}{!}{$L = L_{mlm} + 0.3 * L_{dis} + 0.1 * L_{ss} + 0.03 * L_{clash} + 2*L_{FAPE} + 0.1* L_{ss3d} + 0.01 * L_{pLDDT}.$}
\end{align}

\subsubsection*{Structure relaxation by force fields}
As a preventive measure to resolve any remaining structural clashes and violations, we may relax our model predictions using a restrained energy minimization procedure, such as AMBER \cite{salomon2013overview} and BRiQ \cite{xiong2021pairing}. Specifically, we minimized the AMBER force field using harmonic restraints, allowing the system to maintain a close relationship with its input structure. This post-prediction relaxation also enforces the geometric features of phosphodiester bonds. Our empirical evidence indicates, as measured by RMSD and TM-score, that while this final relaxation does not improve the model's accuracy, it eliminates distracting stereochemical violations without compromising accuracy.

\subsubsection*{Implementation details and running time}

We used the Adam optimizer with a 0.0003 learning rate for 300,000 iterations, alongside a polynomial decay scheduler with 10,000 warm-up steps and a batch size of 16. A dropout ratio of 0.1 was applied to the Rhoformer and structure modules during training. The hardware setup included a GPU cluster with 768 GB memory and eight NVIDIA A100 GPUs (80 GB each), supported by an Intel Xeon Gold 6230 CPU @ 2.10GHz with 64 cores. RhoFold+ was trained for 1,600 epochs over 300,000 iterations, taking approximately one week. Post-training, the inference is rapid, with RhoFold+ predicting a structure in about 0.14 seconds on a single A100 GPU. For FARFAR2 benchmarking, which demands significant computational resources, a Slurm job was run on the cluster using a single CPU core and 8 GB memory, with execution times detailed in Supplementary Table 9.

\subsection*{Running other baselines}
In our benchmarking experiments, we obtained DeepFoldRNA, DRfold, RoseTTAFold2NA, FARFAR2, and trRosettaRNA (v1.0) from their official code repositories (either their homepages or their GitHub repositories). The authors of AlphaFold3 did not publish the code, so we used their server to perform predictions. For FARFAR2, we followed the default settings specified in the corresponding code documentation or on the server, and we trained 100 models. Our benchmarking and evaluation used CASP15 natural RNA targets and newly determined single-stranded RNA structures. 
\highlight{Following CASP15 guidelines, we collected 5 candidate models for each competing method to compute RMSD and Z-score. For AlphaFold3, which generates 5 models per input sequence per run, we conducted one run and collected the 5 models; for RhoFold+, we ran 5 times with different sampled MSAs; and for the other methods, we collected their 5 models from CASP15 website. For the newly determined single-stranded RNA structures, we ran the default configurations of RhoFold+ and other methods to produce default predictions for each sequence. For AlphaFold3, only the top model (‘model\_0’) of a single run was evaluated.
The input of all methods consisted solely of RNA sequences, without ions or other molecules.}

%% file: data.tex
All data used in our work were obtained from related public datasets. We obtained all the RNA 3D structures using the data list arranged by BGSU RNA Representative Sets (version 2022-04-13) (\url{http://rna.bgsu.edu/rna3dhub/nrlist/release/3.226})
and downloaded them from Protein Data Bank (\url{https://www.rcsb.org}).

For pre-training our language model (RNA-FM), we downloaded the unannotated RNA sequences from RNAcentral (\url{https://rnacentral.org/}).
For RNA MSA construction, we built the database using a nucleotide database (\url{ftp://ftp.ncbi.nlm.nih.gov/blast/db/FASTA/nt.gz}), Rfam (\url{https://rfam.xfam.org}) and RNAcentral (\url{https://rnacentral.org}) and use rMSA (\url{https://github.com/pylelab/rMSA}) for searching and construction tools.
We used secondary structural information for self-distillation.
For this data, we downloaded the bpRNA dataset from SPOT-RNA at \url{https://sparks-lab.org/server/spot-rna/}, bprna-1m data from \url{https://bprna.cgrb.oregonstate.edu/} and used RNAStralign, based on E2Efold, from \url{https://github.com/ml4bio/e2efold}. 
The family/type information in Rfam (\url{https://rfam.xfam.org}) was used for cross-family/type validation.
For RNA-Puzzles, we downloaded native structures and submissions of other methods from  \url{https://github.com/RNA-Puzzles/standardized_dataset} and \url{http://www.rnapuzzles.org/results/}, respectively. Similarly, CASP15 data is obtained via  \url{https://predictioncenter.org/casp15/index.cgi}.

%% file: code.tex
For the RhoFold+ model, trained weights, and inference scripts are available under an open-source license at \url{https://github.com/ml4bio/RhoFold}. RhoFold+ is also freely available as a server for academic purposes at \url{https://proj.cse.cuhk.edu.hk/aihlab/RhoFold/#/}.

Our pre-trained language model (RNA-FM) and its inference pipeline can be found at \url{https://github.com/ml4bio/RNA-FM}. The RNA MSA search was performed by combining Infernal (\url{http://eddylab.org/infernal/}), Blastn (\url{https://blast.ncbi.nlm.nih.gov/Blast.cgi}), HMMER (\url{http://hmmer.org}), and rMSA (\url{https://github.com/pylelab/rMSA}), we also used openmm 7.7 for AMBER forcefield relaxation.

Source codes are written under Python 3.7. We also utilized the following software for data collection, data analysis, and visualization: Infernal 1.1.3 (cmbuild, cmcalibrate, cmscan, cmsearch), CD-HIT 4.8.1 (cd-hit-est), HMMER 3.3 (nhmmer), HH-suite 2.0.15, numpy 1.21.2, PyTorch 1.10.2, pandas 1.3.1, matplotlib 3.4, scikit-learn 0.24, scipy 1.7.1, biopython 1.79, PyTorch-Ignite 0.4.6, and TensorBoard 2.6.0.

%% file: acknowledgment.tex
\textbf{Funding:} This work was supported by the Chinese University of Hong Kong (CUHK; award numbers 4937025, 4937026, 5501517, 5501329, 8601603, 8601663) and the Research Grants Council of the Hong Kong Special Administrative Region, China (Hong Kong SAR; project numbers CUHK 24204023, CUHK 14222922, and RGC GRF 2151185). 
Additional support was provided by the Innovation and Technology Commission of the Hong Kong SAR (project no. GHP/065/21SZ to Y. Li).
J.W. was supported by a Hong Kong PhD Fellowship (award no. PF22-73180) from the Research Grants Council of the Hong Kong SAR, China, and an IdeaBooster Fund (project no. IDBF24ENG06) from CUHK.
The work is part of the Antibiotics-AI Project, directed by J.J.C., and supported by the Audacious Project, Flu Lab, LLC, the Sea Grape Foundation, R. Zander, H. Wyss for the Wyss Foundation, and an anonymous donor. 
F.W. was supported by the National Institute of Allergy and Infectious Diseases of the NIH (award no. K25AI168451). 
D.L. acknowledges the startup funding from ASU. 
We thank Xiang-Jun Lu for guidance on inter-helical twist experiments.

\noindent \textbf{Publication:} We thank all our collaborators and  the editorial team at \emph{Nature Methods} for facilitating this publication. This work appears in \emph{Nature Methods} 21, 2287–2298 (2024). \href{https://doi.org/10.1038/s41592-024-02487-0}{doi:10.1038/s41592-024-02487-0}